\documentclass
[superscriptaddress,secnumarabic,amssymb,amsmath,nobibnotes,aps,prd,showkeys,showpacs,nofootinbib,onecolumn]{revtex4}%
\usepackage{graphicx}
\usepackage{subfigure}
\usepackage{epstopdf}
\usepackage{epsf}
\usepackage{bm}
\usepackage{amsmath}
\usepackage{amsfonts}
\usepackage{amssymb}%
\providecommand{\U}[1]{\protect\rule{.1in}{.1in}}

\newcommand{\be}{\begin{equation}}
\newcommand{\ee}{\end{equation}}

\newcommand{\mincir}{\raise
-3.truept\hbox{\rlap{\hbox{$\sim$}}\raise4.truept\hbox{$<$}\ }}
\newcommand{\magcir}{\raise
-3.truept\hbox{\rlap{\hbox{$\sim$}}\raise4.truept\hbox{$>$}\ }}

\begin{document}
\title{Integrability and Cosmological Solutions in Einstein-\ae ther-Weyl theory}
\author{Andronikos Paliathanasis}
\email{anpaliat@phys.uoa.gr}
\affiliation{Institute of Systems Science, Durban University of Technology, Durban 4000,
South Africa}
\author{Genly Leon}
\email{genly.leon@ucn.cl}
\affiliation{Departamento de Matem\'{a}ticas, Universidad Cat\'{o}lica del Norte, Avda.
Angamos 0610, Casilla 1280 Antofagasta, Chile}

\begin{abstract}
We consider a Lorentz violating scalar field cosmological model given
by the modified Einstein-\ae ther theory defined in Weyl integrable geometry. The
existence of exact and analytic solutions is investigated for the case of a
spatially flat Friedmann--Lema\^{\i}tre--Robertson--Walker background space.
We show that the theory admits cosmological solutions of special interests. In
addition, we prove that the cosmological field equations admit the Lewis
invariant as a second conservation law, which indicates the integrability of
the field equations.

\end{abstract}
\date{\today}
\maketitle

\section{Introduction}

\label{sec1}

A plethora of modified or alternative theories to Einstein's gravity
\cite{mod1,mod2} have been proposed during the last years in order to  explain
the cosmological observations. There is a family of theories which violate the
Lorentz symmetry. The main representatives of the Lorentz violating
gravitational theories are the Ho\v{r}ava-Lifshitz theory \cite{hor3} and the
Einstein-\ae ther theory \cite{DJ,DJ2}.

Ho\v{r}ava-Lifshitz gravity is a power-counting renormalizable theory with
consistent ultra-violet behaviour exhibiting an anisotropic Lifshitz scaling
between time and space at the ultra-violet limit \cite{hor3}. On the other
hand, in Einstein-\ae ther theory, the quadratic invariants of the kinematic
quantities of a unitary time-like vector field, which is called \ae ther field,
are introduced in the gravitational Action integral; modifying the
Einstein-Hilbert Action \cite{DJ2}. The Einstein-\ae ther action is the most
general second-order theory which is defined by the spacetime metric and the
\ae ther field involving no more than two derivatives \cite{Carroll:2004ai} (not including total derivatives).
 There are several gravitational and cosmological applications for both
of these theories in the literature, for instance see
\cite{Cai:2009qs,Christodoulakis:2011np,Saridakis:2009bv,Kiritsis:2009sh,Lu:2009em,Nilsson:2018knn,Carruthers:2010ii,Zlosnik:2006zu,bh03,col11,latta1,col112,col113,roum1,ch1,ch2,ea05,ea06}
and references therein.

Scalar fields play a significant role in the explanation of the early
acceleration phase of the universe known as inflation. Lorentz violating scalar
field theories have been studied in Ho\v{r}ava-Lifshitz gravity
\cite{hrs1,hrs2,hrs3,hrs4} and in the Einstein-\ae ther theory
\cite{Kanno:2006ty,DJ,DJ00,Barrow:2012qy,bar2,ae1,ae2}. In \cite{Kanno:2006ty}
it has been proposed an Einstein-\ae ther scalar field model in which the
coupling coefficients of the \ae ther field with  gravity are functions of
the scalar fields. From the latter an interaction between the scalar field and
the \ae ther field it follows. For that model it was found that the inflationary
era is divided into two parts, a Lorentz-violating stage and the standard
slow-roll stage. In the Lorentz-violating stage the universe expands as an
exact de Sitter spacetime, although the inflaton field is rolling down the potential.

In this work we are interest on the existence of exact and analytic solutions
for a Lorentz-violating scalar field cosmological model. We consider the
Einstein-\ae ther theory defined in Weyl integrable geometry \cite{lw1}. In this
specific theory the Action Integral of the Einstein-\ae ther is modified such
that a scalar field coupled to the \ae ther field is introduced in a
geometric way. The global dynamics of the background space were studied in \cite{lw1} for various cosmological models in the absence or in
the presence of matter. It was found that the model provides several cosmological eras in agreement with the cosmological history. In addition, a
Weyl manifold is a conformal manifold equipped with a connection which
preserves the conformal structure and is torsion-free. In Weyl  integrable theory the
geometry is supported by the metric tensor and a connection structure which 
differs from the conformal equivalent metric by a scalar field  \cite{cur0,cur1}.
Cosmological and gravitational applications of the Weyl geometry can be found
for instance in
\cite{Aguila:2014moa,va2,Villanueva:2018kem,va1,va3,va4,va5,va6,vaa9}. The
novelty of the Weyl geometry is that the scalar field in the gravitational
Action integral is introduced by the geometry.

The context of integrability is essential in all areas of physics. A set of
differential equations describing a physical system is said to be integrable
if there exist a sufficient number of invariant functions such that the
dynamical system can be written in algebraic form. When the latter algebraic
system is explicitly solved  the solution of the dynamical system can be expressed in closed-form \cite{al1,al2}. The study of integrability
properties of nonlinear dynamical systems is important, because analytical
techniques can be applied for the better understanding of the physical phenomena.
Although, nowadays we have powerful computers and numerical techniques to
solve nonlinear differential equations, as Arscott discussed on the preface of
his book \cite{Arc} ``[...] fall back on numerical techniques savours somewhat
of breaking a door with a hammer when one could, with a little trouble, find
the key". 

The plan of the paper is as follows. 
In Section \ref{sec2} we present the cosmological model under consideration
which is that of Einstein-\ae ther defined in Weyl integrable geometry assuming a spatially
flat Friedmann--Lema\^{\i}tre--Robertson--Walker \ (FLRW) background space
without any matter source terms. In Section \ref{sec3} we present for the
first time analytic and exact solutions for this cosmological model, we focus
on the existence of exact solutions where the scale factor describes
inflationary models of special interests. We obtain those solutions which are
determined as the general analytic solutions for the corresponding scalar
field potentials. In Section \ref{sec4} we show that this is possible because
the cosmological field equations form an integrable dynamical system, where
the conservation laws are the constraint cosmological equation, i.e. the
modified first Friedmann's equation and the Lewis invariant. The later
invariant  is essential for the study of integrability of time-dependent
classical or quantum systems. We show that the field equations form an
integrable dynamical system for an arbitrary potential function. Finally, in
Section \ref{sec5} we summarize our results and we draw our conclusions.

\section{Einstein-\ae ther-Weyl theory}

\label{sec2}

The Einstein-\ae ther-Weyl gravitational model is an extension of\ the Lorentz
violating Einstein-\ae ther theory in Weyl integrable geometry. It is a scalar
field Lorentz violating theory where there is a coupling between the scalar
field and the \ae ther field. The corresponding gravitational Action integral
has the form of Einstein-\ae ther gravity, thus it is generalized in Weyl
integrable geometry. The latter generalization provides a geometric mechanism
for the introduction of the scalar field into the gravitational theory.

 Weyl geometry is a generalization of Riemannian geometry where 
the metric tensor and the covariant derivative $\left\{  {g}_{\mu\nu}%
,{\nabla}_{\mu}\right\},$ are generalized to $\left\{  \tilde{g}_{\mu\nu}%
,\tilde{\nabla}_{\mu}\right\},$ where $\tilde{\nabla}_{\mu}$ is not defined by the Levi-Civita connection of $g_{\mu \nu}$, but by the
affine connection $\tilde{\Gamma}_{\mu\nu}^{\kappa}\left(  \tilde{g}\right)  $
with the property
\begin{equation}
\tilde{\nabla}_{\kappa}g_{\mu\nu}=\omega_{\kappa}g_{\mu\nu},\label{ww.01}%
\end{equation}
and $\tilde{g}_{\mu\nu}$ is the metric compatible with $\tilde{\nabla}_{\mu}$.
We study Weyl integrable geometry, where the gauge vector field $\omega_{\mu}$ which defines the geometry is a gradient vector field, i.e., it satisfies $\omega_{\mu}=\phi_{,\mu}$ for an scalar field $\phi$. Then, it is defined the new metric tensor  $\tilde{g}_{\mu\nu}=e^{-\phi}g_{\mu\nu}$ as the conformal related metric
compatible with the covariant derivative~$\tilde{\nabla}_{\mu}$, i.e, $\tilde{\nabla}_{\kappa} \tilde{g}_{\mu\nu}=0$. Connections
$\tilde{\Gamma}_{\mu\nu}^{\kappa}$ can be constructed from the Christoffel
symbols~$\Gamma_{\mu\nu}^{\kappa}\left(  g\right)  $ of the metric tensor
$g_{\mu\nu}$ as follows \cite{salim96}:%
\begin{equation}
\tilde{\Gamma}_{\mu\nu}^{\kappa}=\Gamma_{\mu\nu}^{\kappa}-\phi_{,(\mu}%
\delta_{\nu)}^{\kappa}+\frac{1}{2}\phi^{,\kappa}g_{\mu\nu}.\label{ww.02}%
\end{equation}

The gravitational integral of the Einstein-\ae ther-Weyl theory is \cite{lw1}: 
\begin{equation}
S_{AEW}\left(  g_{\mu\nu},\tilde{\Gamma}_{\mu\nu}^{\kappa};u^{\mu}\right)
=S_{W}\left(  g_{\mu\nu},\tilde{\Gamma}_{\mu\nu}^{\kappa}\right)
+S_{AE}\left(   {g}_{\mu\nu},\tilde{\Gamma}_{\mu\nu}^{\kappa};u^{\mu}\right),
\label{ww.03}%
\end{equation}
where $S_{W}\left(  g_{\mu\nu},\tilde{\Gamma}_{\mu\nu}^{\kappa}\right)  $ is
the extension of the Einstein-Hilbert action in Weyl geometry \cite{salim96}:
\begin{equation}
S_{W}\left(  g_{\mu\nu},\tilde{\Gamma}_{\mu\nu}^{\kappa}\right)=\int dx^{4}\sqrt{-g}\left(  \tilde{R}+\xi\left(  \tilde{\nabla}_{\nu
}\left(  \tilde{\nabla}_{\mu}\phi\right)  \right)  g^{\mu\nu}\right),
\label{ww.0a}%
\end{equation}
with $ \tilde{R}$ denoting the Weylian scalar curvature 
\begin{equation}
\tilde{R}=R-\frac{3}{\sqrt{-g}}\left(  g^{\mu\nu}\sqrt{-g}\phi\right)
_{,\mu\nu}+\frac{3}{2}\left(  \tilde{\nabla}_{\mu}\phi\right)  \left(
\tilde{\nabla}_{\nu}\phi\right), \label{ww.005}%
\end{equation}
and $\xi$ is an arbitrary coupling constant.
 $S_{AE}\left(  \tilde{g}_{\mu\nu},\tilde{\Gamma}_{\mu\nu}^{\kappa};u^{\mu}\right)$ is the Action Integral for the \ae ther field $u^{\mu}$ defined in
Weyl geometry, that is: 
\begin{equation}
S_{AE}\left(   {g}_{\mu\nu},\tilde{\Gamma}_{\mu\nu}^{\kappa};u^{\mu}\right)
=\int d^{4}x\sqrt{-\tilde{g}}\left(  \tilde{K}^{\alpha\beta\mu\nu}%
\tilde{\nabla}_{\alpha}u_{\mu}\tilde{\nabla}_{\beta}u_{\nu}+\lambda\left(
\tilde{g}_{\mu\nu}u^{\mu}u^{\nu}+1\right)  \right) , \label{ae.01a}%
\end{equation}
where $\tilde{g}_{\mu\nu}=e^{-\phi}g_{\mu\nu}$ is the conformal related metric
associated with the covariant derivative~$\tilde{\nabla}_{\mu}$. 

Parameters $c_{1},~c_{2},~c_{3}$ and $c_{4}$ are dimensionless constants and
define the coupling between the \ae ther field and the conformal metric through a kinetic term.  Lagrange multiplier $\lambda
~$ensures the unitarity of the \ae ther field, i.e. $\tilde{g}_{\mu\nu}u^{\mu
}u^{\nu}=-1,$ while the fourth-rank tensor~$\tilde{K}^{\alpha\beta\mu\nu}$ is
defined as
\begin{equation}
\tilde{K}^{\alpha\beta\mu\nu}\equiv c_{1}\tilde{g}^{\alpha\beta}\tilde{g}%
^{\mu\nu}+c_{2}\tilde{g}^{\alpha\mu}\tilde{g}^{\beta\nu}+c_{3}\tilde
{g}^{\alpha\nu}\tilde{g}^{\beta\mu}+c_{4}\tilde{g}^{\mu\nu}u^{\alpha}u^{\beta
}.\label{ae.02}%
\end{equation}

Equivalently,
the Action Integral (\ref{ae.01a}) can expressed in terms of the kinematic
quantities $\left\{  \tilde{\theta},\tilde{\sigma}_{\mu\nu},\tilde{\omega
}_{\mu\nu},\tilde{\alpha}^{\mu}\right\}  $ as follows
\begin{equation}
S_{AE}\left(  g_{\mu\nu},\tilde{\Gamma}_{\mu\nu}^{\kappa};u^{\mu}\right)
=\int\sqrt{-\tilde{g}}dx^{4}\left(  \frac{c_{\theta}}{3}\tilde{\theta}%
^{2}+c_{\sigma}\tilde{\sigma}^{2}+c_{\omega}\tilde{\omega}^{2}+c_{\alpha
}\tilde{\alpha}^{2}\right)  ,
\end{equation}
where the new parameters $c_{\theta},~c_{\sigma},~c_{\omega},~c_{a}$ are
functions of $c_{1},~c_{2},~c_{3}$ and $c_{4}$, that is, $c_{\theta}=\left(
c_{1}+3c_{2}+c_{3}\right)  $~,\ $c_{\sigma}=c_{1}+c_{3}~$,$\ c_{\omega}%
=c_{1}-c_{3}~$,$\ c_{a}=c_{4}-c_{1}$.

From (\ref{ww.03}) we find the gravitational field equations to be
\begin{equation}
\tilde{G}_{\mu\nu}+\tilde{\nabla}_{\nu}\left(  \tilde{\nabla}_{\mu}%
\phi\right)  -\left(  2\xi-1\right)  \left(  \tilde{\nabla}_{\mu}\phi\right)
\left(  \tilde{\nabla}_{\nu}\phi\right)  +\xi g_{\mu\nu}g^{\kappa\lambda
}\left(  \tilde{\nabla}_{\kappa}\phi\right)  \left(  \tilde{\nabla}_{\lambda
}\phi\right)  -g_{\mu\nu}U(\phi)  ={T_{\mu\nu}^{\ae },}
\label{ae.022}%
\end{equation}
where ${T_{ab}^{\ae }}$ is the energy momentum tensor which corresponds to the
\ae ther field and $\tilde{G}_{\mu\nu}$ is the Einstein tensor in Weyl theory,
that is, $\tilde{G}_{\mu\nu}=\tilde{R}_{\mu\nu}-\frac{1}{2}\tilde{R}g_{\mu\nu
}$. The rhs of equation (\ref{ae.022}) corresponds to the energy-momentum tensor of the
\ae ther field: %
\begin{align}
{T_{\mu\nu}^{\ae }}  &  =2c_{1}(\tilde{\nabla}_{\mu}u^{\alpha}\tilde{\nabla
}_{\nu}u_{\alpha}-\tilde{\nabla}_{\alpha}u_{\mu}\tilde{\nabla}_{\beta}u_{\nu
}\tilde{g}^{\alpha\beta})+2\lambda u_{\mu}u_{\nu}+\tilde{K}^{\mu\beta}%
{}_{\alpha\mu}\tilde{\nabla}_{\mu}u^{\alpha}\tilde{\nabla}_{\beta}u^{\mu
}g_{\mu\nu}\nonumber\\
&  -2[\tilde{\nabla}_{\alpha}(u_{(\mu}J^{\alpha}{}_{\nu)})+\tilde{\nabla
}_{\alpha}(u^{\alpha}J_{(a\nu)})-\tilde{\nabla}_{\alpha}(u_{(\mu}J_{\nu)}%
{}^{\alpha})]-2c_{4}\left(  \tilde{\nabla}_{\alpha}u_{\mu}u^{\alpha}\right)
\left(  \tilde{\nabla}_{\beta}u_{\nu}u^{\beta}\right)  ,
\end{align}
where ${{J}}{{^{\mu}}_{\nu}}=-\tilde{K}{{^{\mu\beta}}_{\nu\alpha}\tilde
{\nabla}}_{\beta}{u}^{\alpha}.$

\subsection{FLRW spacetime}

In the case of a spatially flat FLRW geometry, with line element
\begin{equation}
ds^{2}=-dt^{2}+a^{2}\left(  t\right)  \left(  dx^{2}+dy^{2}+dz^{2}\right)  ,
\label{ww.05}%
\end{equation}
for the comoving \ae ther field we calculate%
\begin{equation}
\tilde{\theta}=\theta-\dot{\phi}~,~\tilde{\sigma}^{2}=0~,~\tilde{\omega}%
^{2}=0~\text{and }\tilde{\alpha}^{2}=0
\end{equation}
where $\theta$ is the Riemannian expansion rate defined as $\theta=3\frac
{\dot{a}}{a}$. The gravitational field equations are expressed as follows%
\begin{equation}
\frac{\theta^{2}}{3}-\rho_{\phi}-\rho^{\text{\ae \ }}=0 \label{ww.06}%
\end{equation}%
\begin{equation}
\dot{\theta}+\frac{\theta^{2}}{3}+\frac{1}{2}\left(  \rho_{\phi}+3p_{\phi
}\right)  +\frac{1}{2}\left(  \rho^{\text{\ae \ }}+3p^{\text{\ae }}\right)  =0
\label{ww.07}%
\end{equation}
where $\rho_{\phi},$~$p_{\phi}$ are the energy density and pressure of the
scalar field, that is,
\begin{equation}
\rho_{\phi}\left(  \phi,\dot{\phi}\right)  =\frac{\zeta}{2}\dot{\phi}%
^{2}-U(\phi)  ~,~p_{\phi}\left(  \phi,\dot{\phi}\right)
=\frac{\zeta}{2}\dot{\phi}^{2}+U(\phi)  ,
\end{equation}
where parameter $\zeta:= 2 \xi -\frac{3}{2}$ is a coupling parameter between the scalar field and
the gravity. Furthermore, $\rho^{\text{\ae \ }},~p^{\text{\ae }}$ are the density and pressure of the \ae ther field, defined as
\begin{equation}
\rho^{\text{\ae \ }}=-\frac{c_{\theta}}{3}\tilde{\theta}^{2},~p^{\text{\ae \ }%
}=\frac{c_{\theta}}{3}\left(  2\tilde{\theta}_{,t}+\tilde{\theta}^{2}\right)
\end{equation}
in which $h_{\mu\nu}=g_{\mu\nu}+u_{\mu}u_{\nu}$, that is%
\begin{equation}
\rho^{\text{\ae \ }}=-\frac{c_{\theta}}{3}\left(  \theta-\dot{\phi}\right)
^{2},~p^{\text{\ae \ }}=\frac{c_{\theta}}{3}\left(  2\left(  \dot{\theta
}-\ddot{\phi}\right)  +\left(  \theta-\dot{\phi}\right)  ^{2}\right)  \,.
\end{equation}

By replacing in (\ref{ww.06}), (\ref{ww.07}) we derive the gravitational field
equations
\begin{equation}
\left(  1+c_{\theta}\right)  \frac{\theta^{2}}{3}-\frac{2}{3}c_{\theta}%
\theta\dot{\phi}-\left(  \frac{\zeta}{2}-\frac{c_{\theta}}{3}\right)
\dot{\phi}^{2}-U(\phi)  =0, \label{ff.07}%
\end{equation}%
\begin{equation}
\left(  1+c_{\theta}\right)\dot{\theta}+\frac{\left(  1+c_{\theta}\right)  }{3}  \theta^{2}-\frac{2}{3}c_{\theta}\theta\dot{\phi}+\left(  \zeta
+\frac{c_{\theta}}{3}\right)  \dot{\phi}^{2}-c_{\theta}\ddot{\phi}-U\left(
\phi\right)  =0, \label{ff.08}%
\end{equation}

while the equation of motion for the scalar field is
\begin{equation}
\left(  2c_{\theta}-3\left(  1+c_{\theta}\right)  \zeta\right)  \ddot{\phi
}+3\zeta c_{\theta}\dot{\phi}^{2}-3\left(  1+c_{\theta}\right)  \zeta
\theta\dot{\phi}-3\left(  1+c_{\theta}\right)  U_{,\phi}=0. \label{ff.09}%
\end{equation}

In the following we investigate the existence of analytic solutions for the
dynamical system (\ref{ff.07})-(\ref{ff.09}). It is important to mention that
the three equations are not independent, indeed equation (\ref{ff.07}) is a
conservation law for the higher-order equations (\ref{ff.08}), (\ref{ff.09}).

It is important to mention that in the latter dynamical system for $\zeta=0$
the system is degenerated and it has only one dependent variable, hence, in the
following we consider that $\zeta\neq0$. \

\section{Exact solutions}

\label{sec3}

Before to proceed with the derivation of the exact solution we perform the
following change of variable $\phi\left(  t\right)  =-\frac{2}{3}%
\frac{c_{\theta}}{\zeta}\ln\psi\left(  t\right)  $, and define $V(\psi)= U(-\frac{2}{3}%
\frac{c_{\theta}}{\zeta}\ln\psi\left(  t\right))$,
where now the
gravitational field equations becomes%
\begin{equation}
\left(  1+c_{\theta}\right)  \frac{\theta^{2}}{3}+\frac{4}{9}\frac{\left(
c_{\theta}\right)  ^{2}}{\zeta}\frac{\dot{\psi}}{\psi}\theta+\frac{2\left(
c_{\theta}\right)  ^{2}\left(  2c_{\theta}-3\zeta\right)  }{27\zeta^{2}%
}\left(  \frac{\dot{\psi}}{\psi}\right)  ^{2}-V\left(  \psi\right)  =0,
\label{ff.10}%
\end{equation}%
\begin{equation}
\left(  1+c_{\theta}\right)  \left( \dot{\theta}+\frac{1}{3}\theta
^{2}\right)  +\frac{4}{9}\frac{\left(  c_{\theta}\right)  ^{2}}{\zeta}%
\frac{\dot{\psi}}{\psi}\theta+\frac{2\left(  c_{\theta}\right)  ^{2}\left(
2c_{\theta}-3\zeta\right)  }{27\zeta^{2}}\left(  \frac{\dot{\psi}}{\psi
}\right)  ^{2}+\frac{2}{3}\frac{\left(  c_{\theta}\right)  ^{2}}{\zeta}%
\frac{\ddot{\psi}}{\psi}-V\left(  \psi\right)  =0, \label{ff.11}%
\end{equation}%
\begin{equation}
 \frac{4 c_{\theta}^2 (3 (c_{\theta}+1) \zeta -2 c_{\theta})}{27 (c_{\theta}+1) \zeta ^2 }\frac{\ddot{\psi}}{\psi}+\frac{4 c_{\theta}^2 (2 c_{\theta}-3 \zeta )}{27 \zeta ^2}\left(
\frac{\dot{\psi}}{\psi}\right)  ^{2}+\frac{4 c_{\theta}^2 }{9 \zeta  }\frac{\dot{\psi}}{\psi}\theta +\psi V'(\psi)=0. \label{ff.12}%
\end{equation}

Finally, the gravitational equations reduce to
\begin{align}
 & V(\psi)= \frac{2 c_{\theta}^2 (2 c_{\theta}-3 \zeta ) \dot{\psi}^2}{27 \zeta ^2 \psi^2}+\frac{4
   c_{\theta}^2 \theta \dot{\psi}}{9 \zeta  \psi}+\frac{1}{3} (c_{\theta}+1) \theta^2,\\
   & V'(\psi)= \frac{4
   c_{\theta}^2 (3 \zeta -2 c_{\theta}) \dot{\psi}^2}{27 \zeta ^2 \psi^3}-\frac{4 c_{\theta}^2 \theta \dot{\psi}}{9 \zeta  \psi^2}+\frac{2 (3 (c_{\theta}+1) \zeta -2 c_{\theta}) \dot{\theta}}{9 \zeta  \psi}, \\
   & \ddot{\psi}=
   -\frac{3 (c_{\theta}+1) \zeta  \psi \dot{\theta}}{2 c_{\theta}^2}.
   \end{align}

We proceed by study the existence of exact solution for the scalar field for
specific forms of the scalar factor $a\left(  t\right)  $ which describe exact
solutions of special interests.

\subsection{Power-law solution}

Consider the power-law solution $\theta\left(  t\right)  =\frac{2c_{\theta
}^{2}}{3\left(  1+c_{\theta}\right)  \zeta}\frac{\theta_{0}}{t},$ which
describe a universe dominated by an ideal gas with constant equation of state
parameter $w$ and scale factor $a\left(  t\right)  =a_{0}t^{\frac{2}{3\left(
1+w\right)  }}$, with $w=-1+\frac{3\zeta\left(  1+c_{\theta}\right)  }%
{\theta_{0}c_{\theta}^{2}}$, the solution describes an inflationary universe
when $w<-\frac{1}{3}$, while in the special cases where $\theta_{0}%
=\frac{3\zeta\left(  1+c_{\theta}\right)  }{c_{\theta}^{2}},~\theta_{0}%
=\frac{9\zeta\left(  1+c_{\theta}\right)  }{4c_{\theta}^{2}}~$\ or
$\theta_{0}=\frac{3\zeta\left(  1+c_{\theta}\right)  }{2c_{\theta}^{2}}$, the
ideal gas is that of dust fluid, radiation or stiff fluid respectively.

\begin{figure}[ptb]
\centering\includegraphics[width=0.5\textwidth]{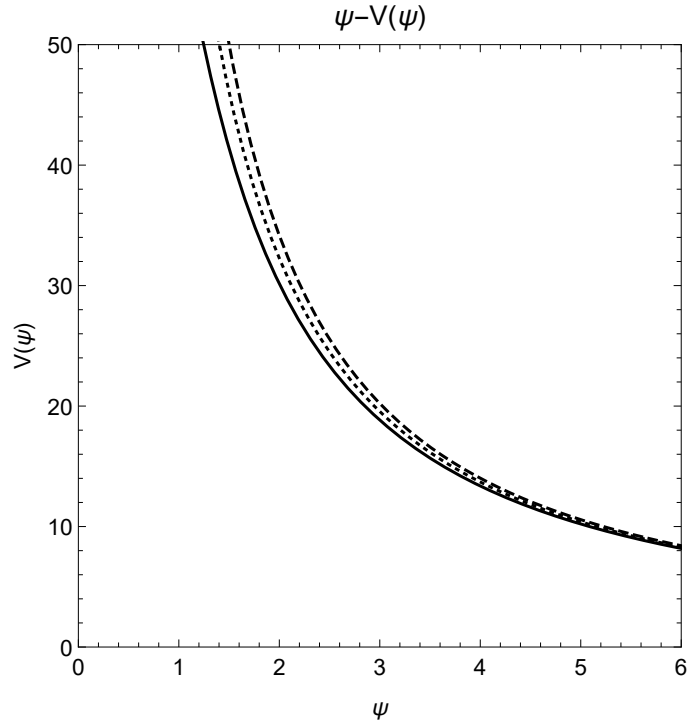} \caption{Qualitative
behaviour of the scalar field potential $V\left(  \psi\right)  $ for various
values of the free parameters $\psi_{1},~\psi_{2}$. \ Solid line is for
$\left(  \psi_{1},\psi_{2}\right)  =\left(  \frac{1}{2},\frac{1}{2}\right)  $,
dotted line is for $\left(  \psi_{1},\psi_{2}\right)  =\left(  \frac{1}%
{2},0\right)  $, dashed line is for $\left(  \psi_{1},\psi_{2}\right)
=\left(  \frac{1}{2},-\frac{1}{2}\right)  $. From the plots we observe that
the potential behaves like a power-law function. The plots are for $\theta
_{0}=1,~\zeta=1$ and $c_{\theta}=6$ where the power-law solution describes an
accelerated universe. The potentials are for the power-law scale factor}%
\label{fig1}%
\end{figure}

By replacing in (\ref{ff.10}) and (\ref{ff.11}) we calculate for the scalar
field
\begin{equation}
\psi\left(  t\right)  =\psi_{1}t^{p_{+}}+\psi_{2}t^{p_{-}}~,~p_{\pm}=\frac
{1}{2}\left(  1\pm\sqrt{1+4\theta_{0}}\right)  , \label{ff.14}%
\end{equation}%
\begin{align}
\frac{27\left(  1+c_{\theta}\right)  \zeta^{2}}{2c_{\theta}^{2}}t^{2}\psi
V\left(  \psi\left(  t\right)  \right)   &  =2\left(  c_{\theta}\right)
^{2}\left(  \psi_{1}\left(  p_{+}+\theta_{0}\right)  t^{p_{+}}+\psi_{2}\left(
p_{-}+\theta_{0}\right)  t^{p_{-}}\right) \nonumber\\
&  +\left(  2c_{\theta}-3\left(  1+c_{\theta}\right)  \zeta\right)  \left(
p_{1}\psi_{1}t^{p_{+}}+p_{2}\psi_{2}t^{p_{-}}\right)  ^{2}. \label{ff.15}%
\end{align}
In Fig. \ref{fig1}, we present the parametric plot for the scalar field
potential $V\left(  \psi\right)  $ as it is expressed by (\ref{ff.15}). In the
special limiting case where $\psi_{1}\psi_{2}=0$, lets say that $\psi_{2}=0$, the
exact solution for the scalar field potential becomes~$V\left(  \psi\right)
=\frac{V_{0}^{A}}{t^{2}}$ with $V_{0}^{A}=\frac{2c_{\theta}^{2}\zeta^{2}\left(
c_{\theta}p_{+}\left(  2-3\zeta\right)  -3p_{+}^{2}\zeta+2c_{\theta}%
^{2}\left(  p_{1}+\theta_{0}\right)  \right)  }{27\left(  1+c_{\theta}\right)
}$; thus, we end up with the power-law potential
\begin{equation}
V\left(  \psi\right)  =V_{0}^{A}\left(  \psi_{1}\right)  ^{\frac{2}{p_{+}}%
}\psi^{-\frac{2}{p_{+}}}. \label{ff.16}%
\end{equation}

\subsection{de Sitter spacetime}

We assume now that the expansion rate $\theta\left(  t\right)  $ is a
constant, i.e. $\theta\left(  t\right)  =\theta_{0}$. That solution describes
the de Sitter universe with scale factor $a\left(  t\right)  =a_{0}%
e^{\frac{\theta_{0}}{3}t}$. Hence from (\ref{ff.10}) and (\ref{ff.11}) we find
the closed-form solution for the scalar field
\begin{equation}
\psi\left(  t\right)  =\psi_{1}\left(  t-t_{0}\right)  , \label{ff.17}%
\end{equation}
where $\psi_{1},~t_{0}$ are integration constants; for the scalar field
potential it follows%
\begin{equation}
V\left(  \psi\left(  t\right)  \right)  =\frac{\left(  1+c_{\theta}\right)
}{3}\theta_{0}^{2}+\frac{4c_{\theta}^{2}}{9\zeta}\frac{\theta_{0}}{\left(
t-t_{0}\right)  }+\frac{2\left(  c_{\theta}\right)  ^{2}\left(  2c_{\theta
}-3\zeta\right)  }{27\zeta^{2}\left(  t-t_{0}\right)  ^{2}}. \label{ff.18}%
\end{equation}
Finally we end with the functional form of the potential%
\begin{equation}
V\left(  \psi\left(  t\right)  \right)  =V_{0}^{B}+V_{1}^{B}\psi^{-1}%
+V_{1}^{C}\psi^{-2}. \label{ff.19}%
\end{equation}
where $V_{0}^{B}=\frac{\left(  1+c_{\theta}\right)  }{3}\theta_{0}^{2}%
,~V_{1}^{B}=\frac{4c_{\theta}^{2}}{9\zeta}\psi_{1}\theta_{0}$ and $V_{1}%
^{C}=\frac{2\left(  c_{\theta}\right)  ^{2}\left(  2c_{\theta}-3\zeta\right)
}{27\zeta^{2}}\left(  \psi_{1}\right)  ^{2}$.

\subsection{Quadratic Lagrangian inflation}

Let us consider the scale factor $a\left(  t\right)  =a_{0}\exp\left(
-a_{1}t^{2}\right)  $ which describes an exact solution of Einstein's General
Relativity with quadratic corrections terms \cite{qq}. This solution can also
be recovered by a modified Chaplygin gas in General Relativity \cite{anin}.
For this scale factor we calculate $\theta\left(  t\right)  =-\frac
{2\theta_{0}}{3\left(  1+c_{\theta}\right)  \zeta}t$,~where we have set
$a_{1}=-\frac{\theta_{0}c_{\theta}^{2}}{9\left(  1+c_{\theta}\right)  \zeta}$.

Therefore, from the field equations (\ref{ff.10}) and (\ref{ff.11}) it follows%
\begin{equation}
\psi\left(  t\right)  =\psi_{1}e^{\sqrt{\theta_{0}}t}+\psi_{2}e^{-\sqrt
{\theta_{0}}t}, \label{ff.20}%
\end{equation}%
\begin{align}
\frac{27\left(  1+c_{\theta}\right)  \zeta^{2}}{2\theta_{0}c_{\theta}^{2}%
}\left(  e^{2\sqrt{\theta_{0}}t}\psi_{1}+\psi_{2}\right)  ^{2}V\left(
\psi\left(  t\right)  \right)   &  =2c_{\theta}^{2}\left(  e^{2\sqrt
{\theta_{0}}t}\left(  \sqrt{\theta_{0}}t-1\right)  \psi_{1}+\psi_{2}\left(
\sqrt{\theta_{0}t}+1\right)  \right)  ^{2} \nonumber\\
&  +\left(  3\left(  1+c_{\theta}\right)  \zeta-2c_{\theta}\right)  \left(
e^{2\sqrt{\theta_{0}}t}\psi_{1}-\psi_{2}\right)  ^{2}. \label{ff.21}%
\end{align}

For $\psi_{2}=0$, the scalar field potential is written as%
\begin{equation}
V\left(  \psi\right)  =V_{0}^{C}+V_{1}^{C}\ln\left(  \frac{\psi}{\psi_{1}%
}\right)  +V_{2}^{C}\left(  \ln\left(  \frac{\psi}{\psi_{1}}\right)  \right)
^{2}, \label{ff.22}%
\end{equation}
where $V_{0}^{C}=\frac{2\theta_{0}\left(  2c_{\theta}\left(  1+c_{\theta
}\right)  -3\zeta\right)  }{27\left(  1+c_{\theta}\right)  \zeta^{2}}$%
,$~V_{1}^{C}=-\frac{8\theta_{0}c_{\theta}^{4}}{27\left(  1+c_{\theta}\right)
\zeta^{2}}$ and $V_{2}^{C}=\frac{4\theta_{0}c_{\theta}^{4}}{27\left(
1+c_{\theta}\right)  \zeta^{2}}$.

\subsection{Scale factor $a\left(  t\right)  =a_{0}t^{\alpha_{1}}e^{\alpha
_{2}t}$}

Scale factor of the form $a\left(  t\right)  =a_{0}t^{\alpha_{1}}e^{\alpha
_{2}t}~$has been studied before in \cite{anin}. For this solution we find
$\theta\left(  t\right)  =3\left(  \frac{a_{1}}{t}+a_{2}\right)  $, and
$\dot{\theta}\left(  t\right)  =3\frac{a_{1}}{t}$. For simplicity we replace
$a_{1}=\frac{\theta_{0}c_{\theta}^{2}}{9\zeta\left(  1+c_{\theta}\right)  }$,
while for the scalar field it follows%
\begin{equation}
\psi\left(  t\right)  =\psi_{1}t^{q_{+}}+\psi_{2}t^{q_{-}}~,~q_{\pm}=\frac
{1}{2}\left(  1+\sqrt{1+2\theta_{0}}\right)  ,
\end{equation}%
\begin{align}
27t^{2}\zeta^{2}V\left(  \psi\left(  t\right)  \right)   &  =4c_{\theta}%
^{3}q_{-}^{2}-6c_{\theta}^{2}q_{-}\left(  q_{-}-6a_{2}t\right)  \zeta
+81a_{2}\left(  1+c_{\theta}\right)  t^{2}\zeta^{2}\nonumber\\
&  +\frac{2c_{\theta}^{2}\left(  2c_{\theta}^{2}q_{-}+9a_{2}\left(
1+c_{\theta}\right)  t\zeta\right)  }{1+c_{\theta}}\theta_{0}+\frac{c_{\theta
}^{4}}{1+c_{\theta}}+\nonumber\\
&  +\frac{4c_{\theta}^{2}\left(  q_{+}-q_{-}\right)  t^{q_{+}}\left(  \left(
1+c_{\theta}\right)  \left(  2c_{\theta}q_{-}-3q_{-}\zeta+9a_{2}t\zeta\right)
+\theta_{0}c_{\theta}^{2}\right)  }{\left(  1+c_{\theta}\right)  \psi\left(
t\right)  }\psi_{1}+\nonumber\\
&  +\frac{2c_{\theta}^{2}\left(  q_{+}-q_{-}\right)  t^{2q_{1}}\left(
2c_{\theta}-3\zeta\right)  }{\psi\left(  t\right)  ^{2}}\psi_{1}^{2}.
\end{align}

\begin{figure}[ptb]
\centering\includegraphics[width=0.5\textwidth]{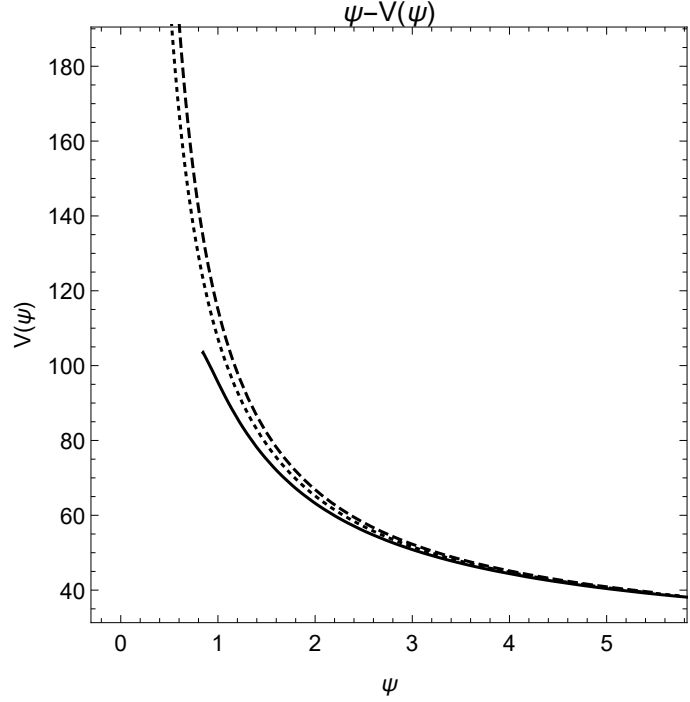} \caption{Qualitative
behaviour of the scalar field potential $V\left(  \psi\right)  $ for various
values of the free parameters $\psi_{1},~\psi_{2}$. \ Solid line is for
$\left(  \psi_{1},\psi_{2}\right)  =\left(  \frac{1}{2},\frac{1}{2}\right)  $,
dotted line is for $\left(  \psi_{1},\psi_{2}\right)  =\left(  \frac{1}%
{2},0\right)  $, dashed line is for $\left(  \psi_{1},\psi_{2}\right)
=\left(  \frac{1}{2},-\frac{1}{2}\right)  $. From the plots we observe that
the potential behaves like a power-law function. The plots are for $\theta
_{0}=1,~\zeta=1$ and $c_{\theta}=6$ and $a_{2}=1.~$The potential is for the
scale factor $a\left(  t\right)  =a_{0}t^{\alpha_{1}}e^{\alpha_{2}t}.$}%
\label{fig2}%
\end{figure}

For $\psi_{2}=0,$ it follows%
\begin{equation}
U(\phi)  =V_{0}^{D}+V_{1}^{D}\left(  \frac{\psi}{\psi_{1}%
}\right)  ^{-\frac{1}{q_{+}}}+V_{2}^{D}\left(  \frac{\psi}{\psi_{2}}\right)
^{-\frac{2}{q_{+}}},
\end{equation}
where $V_{0}^{D}=3a_{2}^{2}\left(  1+c_{\theta}\right)  $, $V_{1}^{D}%
=\frac{2a_{2}c_{\theta}^{2}\left(  2q_{+}+\theta_{0}\right)  }{3\zeta}~$and
$V_{2}^{D}=\frac{2c_{\theta}^{2}\left(  1+c_{\theta}\right)  q_{+}^{2}\left(
2c_{\theta}-3\zeta\right)  +4c_{\theta}^{4}q_{+}\theta_{0}+c_{\theta}%
^{4}\theta_{0}^{2}}{27\left(  1+c_{\theta}\right)  \zeta^{2}}$. \ The
parametric plot of the scalar field potential $V\left(  \psi\right)  $ is
presented in Fig. \ref{fig2} for various values of the free parameters.

\subsection{Intermediate inflation}
\begin{figure}[ptb]
\centering\includegraphics[width=0.5\textwidth]{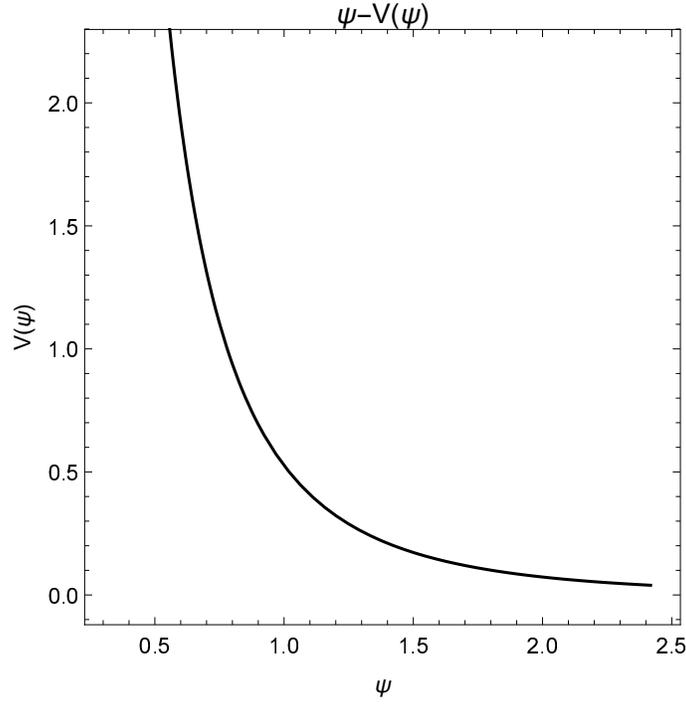} \caption{Qualitative
behaviour of the scalar field potential $V\left(  \psi\right)$ for the intermediated inflation model, plot is for $(\psi_{1},\psi_{2},c_{\theta},\zeta,f,A)=(0,2,1,-1,1/2,1)$.}%
\label{fig5}%
\end{figure}

Consider the power-law solution $\theta\left(  t\right)  =\frac{1}{3}A f t^{-(1-f)}, A>0, 0<f<1$ which describes intermediated inflation with deceleration parameter $q=-1+\frac{(1-f)}{A f} t^{-f}$ and scale factor $a\left(  t\right)  =a_{0} e^{A t^f}$. The solution describes an inflationary universe \cite{if1,if2,if3}. The expansion of the universe with this
scale factor is slower than the de Sitter inflation ($a\left(  t\right)  =a_{0}%
e^{\frac{\theta_{0}}{3}t}$), but
faster than the power law inflation ($a(t)= a_0 t^q$ where $q > 1$). It was shown
that the intermediate inflation arises as the slow-roll solution to potentials which fall off
asymptotically as an inverse power law inflation in the standard canonical framework and 
can be modelled by an exact cosmological solution \cite{if2,if3}. The intermediate inflation has
also been studied in some warm inflationary scenarios in order to examine its predictions
for inflationary observables \cite{if4,if5,if6,Herrera:2018ker}.
With these assumptions the field equations
\eqref{ff.10}, \eqref{ff.11}, \eqref{ff.12}
becomes: 
\begin{align}
 & V(\psi(t))=\frac{1}{27} A^2 (c_{\theta}+1) f^2 t^{2 f-2}+\frac{4 A c_{\theta}^2 f t^{f-1} \dot{\psi}}{27 \zeta  \psi
   (t)}+\frac{2 c_{\theta}^2 (2 c_{\theta}-3 \zeta ) \dot{\psi}^2}{27 \zeta ^2 \psi^2},\\
   & V'(\psi)=-\frac{4 A
   c_{\theta}^2 f t^{f-1} \dot{\psi}}{27 \zeta  \psi^2}+\frac{2 A (f-1) f (3 (c_{\theta}+1) \zeta -2 c_{\theta})
   t^{f-2}}{27 \zeta  \psi}+\frac{4 c_{\theta}^2 (3 \zeta -2 c_{\theta}) \dot{\psi}^2}{27 \zeta ^2 \psi^3}, \\
   & \ddot{\psi}=-\frac{A (c_{\theta}+1) (f-1) f \zeta  t^{f-2} \psi}{2 c_{\theta}^2}.
\end{align}
Choosing $0<f<1, A>0$, we obtain the exact solution
   \begin{align}
& \psi= 2^{-\frac{1}{2 f}} \sqrt{t} (c_{\theta} f)^{-1/f} \left(-A^2 (f-1) f \zeta \right)^{\frac{1}{2 f}} \left|
   c_{\theta}+1\right| ^{\frac{1}{2 f}} \nonumber\\
   & \times \Bigg[\psi_1 \Gamma \left(\frac{f-1}{f}\right) I_{-\frac{1}{f}}\left(\frac{\sqrt{2} A t^{f/2} \sqrt{-(f-1) f \zeta } \sqrt{\left|
   c_{\theta}+1\right| }}{c_{\theta} f}\right) \nonumber \\
   & + \psi_2 (-1)^{\frac{1}{f}} \Gamma \left(1+\frac{1}{f}\right)
   I_{\frac{1}{f}}\left(\frac{\sqrt{2} A t^{f/2} \sqrt{-(f-1) f \zeta } \sqrt{\left| c_{\theta}+1\right| }}{c_{\theta}
   f}\right)\Bigg] ,
   \end{align}
   where  $\psi_1$, $\psi_2$ are integration constants and $I_{1/f}(t)$ denotes the Bessel function. 
   Considering the condition $0<f<1$ we set $\psi_1=0$ to obtain real solutions. In Fig. \ref{fig5} the qualitative behaviour of the latter scalar field potential is presented. 
   
\subsection{Log-mediate inflation}

Consider the solution $\theta\left(  t\right)  =\frac{A \lambda  \ln ^{\lambda -1}(t)}{3 t}$, where $\lambda$ and $A$ are dimensionless constant parameters such that $\lambda> 1$ and $A > 0$,  with deceleration parameter 
$q= -1+ \frac{\ln ^{1-\lambda }(t)}{A \lambda }-\frac{(\lambda -1) \ln ^{-\lambda }(t)}{A
   \lambda }$ and scale factor $a(t) =\exp[A(\ln t)^\lambda]$. 
This generalized model of the expansion of the universe is called
log-mediate inflation \cite{if3,Herrera:2018ker}. Note that for the special case in which $\lambda=1, A=p$, the log-mediate inflation model becomes a power-law inflation model \cite{log1}.

With these assumptions the field equations
\eqref{ff.10}, \eqref{ff.11}, \eqref{ff.12}
becomes: 
\begin{align}
    & V(\psi(t))=\frac{A^2 (c_{\theta}+1) \lambda ^2 \ln ^{2 \lambda -2}(t)}{27 t^2}+\frac{4 A c_{\theta}^2 \lambda  \ln
   ^{\lambda -1}(t) \dot{\psi}}{27 \zeta  t \psi}+\frac{2 c_{\theta}^2 (2 c_{\theta}-3 \zeta ) \dot{\psi}^2}{27 \zeta ^2
   \psi^2},\\
   & V'(\psi(t))=-\frac{4 A c_{\theta}^2 \lambda  \ln ^{\lambda -1}(t) \dot{\psi}}{27 \zeta  t \psi^2}+\frac{2 A
   \lambda  (3 (c_{\theta}+1) \zeta -2 c_{\theta}) (\lambda -\ln (t)-1) \ln ^{\lambda -2}(t)}{27 \zeta  t^2 \psi}+\frac{4 c_{\theta}^2 (3 \zeta -2 c_{\theta}) \dot{\psi}^2}{27 \zeta ^2 \psi^3},\\
   & \ddot{\psi}=\frac{A (c_{\theta}+1) \zeta  \lambda  \psi \ln ^{\lambda -2}(t) (-\lambda +\ln (t)+1)}{2 c_{\theta}^2 t^2}. \label{eq.43}
\end{align}
However, equation \eqref{eq.43} is not integrable in closed form.    
We propose the asymptotic expansion 
\begin{equation}
   \psi(t)\sim c_0 t^{\alpha } \ln (t)+\epsilon  \left(\psi_2 t+\psi_1\right) t^{\alpha -\frac{1}{2}}+O\left(\epsilon ^2\right), \;\;  \text{for} \; t \epsilon < B\;\; \text{for some}\; B>0, \;\; \text{and}\; 0<\epsilon \ll 1.
\end{equation}
The equation \eqref{eq.43} becomes
\begin{align}
 & 0= R(t, \alpha; \epsilon, \lambda, \zeta):=  \frac{c_0 t^{\alpha -2} \left(A \zeta  \lambda  (\lambda -\ln (t)-1) \ln ^{\lambda }(t)+\ln (t) (2 \alpha +(\alpha -1) \alpha  \ln
   (t)-1)\right)}{\ln (t)} \nonumber \\
   & +\epsilon  t^{\alpha -\frac{5}{2}} \left(A \zeta  \lambda  \left(\psi_2 t+\psi_1\right) (\lambda -\ln (t)-1) \ln
   ^{\lambda -2}(t)+\frac{1}{4} (2 \alpha -1) \left((2 \alpha -3) \psi_1+(2 \alpha +1) \psi_2 t\right)\right) +O\left(\epsilon ^2\right),
\end{align}
For $\alpha\in\left\{1, \frac{1}{2}, 0, -\frac{1}{2}, -\frac{5}{2}, \frac{1}{2}\left(1-\sqrt{1+4\lambda}\right)\right\}$, it is verified 
$$\lim_{t\rightarrow \infty } R(t, \alpha; \epsilon, \lambda, \zeta)=0.$$

Moreover, setting $\alpha=\frac{3}{2}-k, k>0$, it follows
\begin{align}
  & R(t, \alpha; \epsilon, \lambda, \zeta):=   \frac{c_0 t^{-k-\frac{1}{2}} \left(4 A \zeta  \lambda  (\lambda -\ln (t)-1) \ln ^{\lambda }(t)+\ln (t) ((4 (k-2) k+3) \ln (t)-8
   k+8)\right)}{4 \ln (t)} \nonumber \\
   & +\frac{\epsilon  t^{-k-1} \left(A \zeta  \lambda  \left(\psi_2 t+\psi_1\right) (\lambda -\ln (t)-1) \ln ^{\lambda
   }(t)+(k-1) \ln ^2(t) \left(\psi_2 (k-2) t+\psi_1 k\right)\right)}{\ln ^2(t)} +O\left(\epsilon ^2\right), \\
  & \lim_{t\rightarrow \infty} R(t, \alpha; \epsilon, \lambda, \zeta)=0. 
\end{align}

Hence, the gravitational field equations in
Einstein-\ae ther-Weyl theory in a spatially flat FLRW background space
described by the set of differential equations (\ref{ff.07})-(\ref{ff.08}) with scale factor $a(t) =\exp[A(\ln t)^\lambda]$ admits an asymptotic solution \begin{align}
  & \psi(t)\sim c_0 t^{\alpha } \ln (t), \\
  & V(\psi(t))\sim \frac{A^2 (c_{\theta}+1) \zeta ^2 \lambda ^2 \ln ^{2 \lambda }(t)+4 A c_{\theta}^2 \zeta  \lambda  (\alpha  \ln (t)+1)
   \ln ^{\lambda }(t)+2 (2 c_{\theta}-3 \zeta ) (\alpha  c_{\theta} \ln (t)+c_{\theta})^2}{27 \zeta ^2 t^2 \ln
   ^2(t)},
\end{align} as $t\rightarrow \infty$, for any $\alpha<\frac{3}{2}$.

That, is
\begin{align}
  & V\sim   O\left(\left(\frac{1}{t}\right)^2\right) \left(\ln ^{2 \lambda }(t)+\ln ^{\lambda }(t)+1\right), \\
  & \psi\sim t^{\alpha } \left(-c_0 \ln
   \left(\frac{1}{t}\right)+O\left(\left(\frac{1}{t}\right)^2\right)\right).
\end{align}

\subsection{$\Lambda$CDM universe}

\begin{figure}[ptb]
\centering\includegraphics[width=0.5\textwidth]{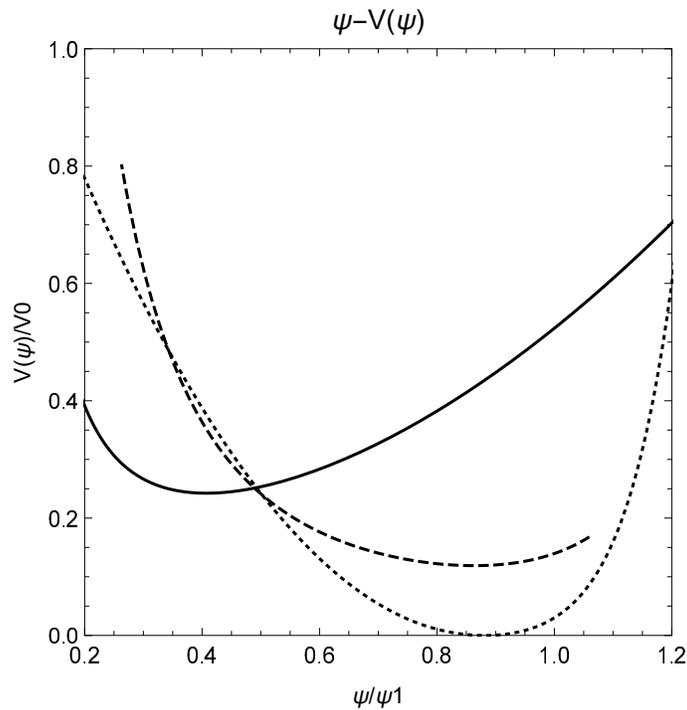} \caption{Qualitative
behaviour of the scalar field potential $V\left(  \psi\right)  $ for various
values of the free parameters $\psi_{1},~\psi_{2}$. \ Solid line is for
$\left(  \psi_{1},\psi_{2}\right)  =\left(  \frac{1}{2},\frac{1}{2}\right)  $,
dotted line is for $\left(  \psi_{1},\psi_{2}\right)  =\left(  \frac{1}%
{2},0\right)  $, dashed line is for $\left(  \psi_{1},\psi_{2}\right)
=\left(  \frac{1}{2},-\frac{1}{2}\right)  $. From the plots we observe that
the potential behaves like a power-law function. The plots are for
$\omega=1,~\zeta=1$ and $c_{\theta}=20$. The axes has been normalized. The
potential is for the $\Lambda$CDM scale factor.}%
\label{fig3}%
\end{figure}

As a final application we consider the scale factor which describes the
$\Lambda$-cosmology, i.e. $a\left(  t\right)  =a_{0}\sinh^{\frac{2}{3}}\left(
\omega t\right)  $. Thus, for this exact solution the scalar field is found to
be expressed in terms of the hypergeometric function
\begin{align}
\psi\left(  t\right)   &  =\psi_{1}\left(  \tanh\left(  \omega t\right)
\right)  ^{\frac{1}{2}-\mu}~_{2}F_{1}\left(  \frac{1}{4}-\frac{\mu}{2}%
,\frac{3}{4}-\frac{\mu}{2},1-\mu,\tanh^{2}\left(  \omega t\right)  \right)
+\\
&  +\psi_{2}\left(  \tanh\left(  \omega t\right)  \right)  ^{\frac{1}{2}+\mu
}~_{2}F_{1}\left(  \frac{1}{4}+\frac{\mu}{2},\frac{3}{4}+\frac{\mu}{2}%
,1+\mu,\tanh^{2}\left(  \omega t\right)  \right)  ,
\end{align}
where $\mu=\frac{\sqrt{c_{\theta}^{2}+12\zeta\left(  1+c_{\theta}\right)  }%
}{2c_{\theta}}$, and $\psi_{1},~\psi_{2}$ are two integration constants. For
simplicity we omit the presentation of the exact form of the scalar field
potential $V\left(  \psi\right)  $. Thus for specific values of the free
parameters we present the parametric evolution of $V\left(  \psi\right)  $ in
Fig. \ref{fig3}.

\section{Integrability of the gravitational field equations}

\label{sec4}

In Section \ref{sec3} we solved the gravitational field equations for different
scale factors, which are of interests as cosmological solutions. The exact
solutions of our analysis have the sufficient number of initial constants of
integration, they are the constants $\psi_{1},~\psi_{2}$ and the
non-essential constant of the time translation $t\rightarrow t+t_{1}~$which we
have omitted. Hence, the solutions that we have found are the general analytic
solutions of the nonlinear dynamical system which provide these specific scale
factors. Note, that we have not considered any functional form for the scalar
field potential but for all the cases that we have studied, a scalar field potential
can be found. Our analysis is motivated by the original work on cosmological
solutions in scalar field theory by Ellis and Madsen \cite{el1}. There, the
solutions that have been found are exact solutions and particularly, they are
special solutions and not the complete solution of the dynamical system. Some
analytic solutions in scalar field cosmology can be found by using techniques of
analytic mechanics such is the theory of invariant transformations
\cite{ans1A,ans2A,ans3A}. However in this study we have not applied any symmetry in
order to find the solutions, that indicates that except from the constraint
equation another conservation law should  always exists for any functional form
of the scalar field potential.

The new scalar field $\psi\left(  t\right)  =\exp\left(  -\frac{3\zeta
}{2c_{\theta}}\phi\left(  t\right)  \right)  $ that we defined it was not an
ad hoc selection. Indeed, in these coordinates by replacing $V\left(
\psi\right)  $ from (\ref{ff.10}) in (\ref{ff.11}) we end with the
second-order differential equation of the form%
\begin{equation}
\ddot{\psi}+\omega\left(  t\right)  \psi=0. \label{e1}%
\end{equation}
where $\omega\left(  t\right)  =\left(  \frac{3}{2}\frac{\left(  1+c_{\theta
}\right)  }{c_{\theta}^{2}}\zeta\dot{\theta}\right)  $.

The second-order differential equation is a linear equation also known as the
time-dependent oscillator \cite{lt1}. The differential equation (\ref{e1})
admits the conservation law \cite{lt2}%
\begin{equation}
I=\frac{1}{2}\left(  \left(  y\dot{\psi}-\dot{y}\psi\right)  ^{2}+\left(
\frac{\psi}{y}\right)  ^{2}\right)  , \label{e2}%
\end{equation}
where $\,y=y\left(  t\right)  $ is any solution of the Ermakov-Pinney
equation
\begin{equation}
\ddot{y}+\omega\left(  t\right)  y-y^{-3}=0. \label{e3}%
\end{equation}

Conservation law (\ref{e2}) it is known as Lewis invariant and it was derived
for the first time as an adiabatic invariant \cite{lt3}. Alternatively, the
conservation law (\ref{e2}) can be constructed through a set of canonical
transformations \cite{lt4} or with the use of Noether's theorem \cite{lt1}.
The set of equations (\ref{e1})-(\ref{e3}) it is also known as the Ermakov
system which can be found in many applications in physical science
\cite{lt5,lt6,lt7,lt8}.

Hence, for the gravitational field equations (\ref{ff.08})-(\ref{ff.09}) the
following theorem holds.

\textbf{Theorem:}\textit{ The gravitational field equations in
Einstein-\ae ther-Weyl theory in a spatially flat FLRW background space
described by the set of differential equations (\ref{ff.07})-(\ref{ff.08})
form an integrable dynamical system for arbitrary potential. The two
conservation laws are the constraint equation (\ref{ff.07}) and the Lewis
invariant}%
\begin{equation}
I\left(  \phi,\dot{\phi},y\right)  =\frac{1}{2}e^{-\frac{3\zeta}{c_{\theta}%
}\phi}\left(  \left(  \frac{3\zeta}{2c_{\theta}}y\dot{\phi}+\dot{y}\right)
^{2}+y^{-2}\right)  ,~
\end{equation}
\textit{where }$y\left(  t\right)  $\textit{ satisfies the Ermakov-Pinney
equation (\ref{e3}).}

It is important to mention at this point that in another lapse function
$dt=N\left(  \tau\right)  d\tau$ in the metric tensor (\ref{ww.05}) our
results are valid. In such a case, the equivalent equation (\ref{e1}) it is of
the form%
\begin{equation}
\frac{d^{2}\psi}{d\tau^{2}}+\alpha\left(  \tau\right)  \frac{d\psi}{d\tau
}+\beta\left(  \tau\right)  \psi=0, \label{e4}%
\end{equation}
which also admits an invariant function \cite{lt4} similar to the Lewis invariant.

Except from the Lewis invariant, the linear differential equation (\ref{e4})
is maximally symmetric and admits eight Lie point symmetries which form the
$SL\left(  3,R\right)  $ Lie algebra \cite{lt10}, for arbitrary functions
$\alpha\left(  \tau\right)  $ and $\beta\left(  t\right)  $. Hence, according
to S. Lie theorem the differential equation (\ref{e4}) is equivalent to the
free particle $Y^{\prime\prime}=0$ and there exists a point transformation
$\left\{  \tau,\psi\left(  \tau\right)  \right\}  \rightarrow\left\{
\chi,Y\left(  \chi\right)  \right\}  $ which transform equation (\ref{e4})
into that of the free particle, for more details we refer the reader to the
review article \cite{moy}. That is an alternative way to prove the
integrability of the gravitational field equations for the cosmological model
of our consideration.

\section{Conclusions}

\label{sec5}

In this work we considered a spatially flat FLRW background space in
Einstein-\ae ther theory defined in Weyl integrable geometry. The novelty of
this approach is that a scalar field coupled to the \ae ther field is introduced
in a geometric way. For this model we investigated the existence of exact
solutions of special interests, in particular we focused on exact solutions
which can describe the inflationary epoch of our universe.

Indeed, we proved that the cosmological model of our consideration can provide
exact solutions such that the power-law inflation, de Sitter expansion,
quadratic Lagrangian inflation and others. For these specific scale factors we
were able to calculate the closed-form expression of the scalar field solution
and of the scalar field potential.

Moreover, we investigate also the possibility of Einstein-\ae ther-Weyl
cosmological model to admit a cosmological solution where the scalar field
unify the dark matter and the dark energy of the universe, and for that
investigation we proved that there exists a scalar field potential which can
describe explicitly the $\Lambda$CDM universe. Scalar field models which unify
the dark components of the universe have been drawn the attention of the
academic society because they provide a simple mechanism for the observable
universe, see \cite{uni1,uni2,uni3,uni4,uni5} and references therein.

However, the main result of this work is that we were able to prove the
integrability of the field equations of our cosmological model for arbitrary
potential function. In particular we found a point transformation which reduce
one of the two equations to the linear equation of the time-dependent
oscillator, and to prove that the Lewis invariant is a conservation law for
the field equations for arbitrary scalar field potential. This is an
interesting result which we did not expect it, assuming the nonlinearity form
of the field equations and mainly that according to our knowledge there is not
any effective Lagrangian description for the cosmological field equations in
order to apply techniques for the investigation of conservation laws similar
with that applied before for the quintessence or the scalar tensor theories.

From this work it is clear that in the background space the
Einstein-\ae ther-Weyl cosmological model is cosmological viable. Thus in a
future work we plan to investigate further the physical properties of this
theory as an inflationary model and as a unified model for the dark components
of the universe.

\begin{acknowledgments}
AP \& GL were funded by Agencia Nacional de Investigaci\'{o}n y Desarrollo -
ANID through the program FONDECYT Iniciaci\'{o}n grant no. 11180126.
Additionally, GL is supported by Vicerrector\'{\i}a de Investigaci\'{o}n y
Desarrollo Tecnol\'{o}gico at Universidad Catolica del Norte.
\end{acknowledgments}

\end{document}